\documentstyle[12pt]{article}

\newtheorem{theorem}{Theorem}
\newtheorem{proposition}{Proposition}
\newtheorem{corollary}{Corollary}

\newcommand{\blackslug}{\mbox{\hskip 1pt \vrule width 4pt height 8pt 
depth 1.5pt \hskip 1pt}}
\newcommand{\QED}{\quad\blackslug\lower 8.5pt\null\par}

\newcommand{\comment}[1]{}

\newcommand{\Bm} {\begin{math}}
\newcommand{\Em} {\end{math}}

\newcommand{\denselist}{\itemsep 0pt} 

\title{
 Bundling Equilibrium in Combinatorial Auctions\footnote{First version: June 2001.}}

\author{
     \begin{tabular}{ccc}
         \large Ron Holzman\thanks{Research supported
         by the Fund for the Promotion of Research at
         the Technion and by Technion V.P.R. Fund -
         M. and M. L. Bank Mathematics Research Fund}
         & \hspace{.2in} & \large Noa Kfir-Dahav\\
  \small Department of Mathematics  & &
 \small  Faculty of IE and Management\\
 \small Technion--Israel Institute of Technology  & &
 \small Technion--Israel Institute of Technology\\
 \small Haifa 32000, Israel  & &
 \small Haifa 32000, Israel\\
  \small holzman@tx.technion.ac.il & &
  \small kfir@ie.technion.ac.il\\
   \small  & & \\
   \small  & & \\
         \large Dov Monderer\thanks{Research supported
         by the Fund for the Promotion of Research at
         the Technion, and by the Israeli Academy of Science.} & \hspace{.2in} & \large Moshe Tennenholtz\\
 \small  Faculty of IE and Management & &
 \small  Computer Science Department\\
 \small Technion--Israel Institute of Technology  & &
 \small Stanford University\\
 \small Haifa 32000, Israel  & &
 \small Stanford, CA 94305, USA\\
  \small dov@ie.technion.ac.il & &
   \small moshe@robotics.stanford.edu\\
     \end{tabular}
  }

\date{September 18, 2001}

\begin{document}

\maketitle

\newpage

\noindent{\bf Abstract:}

This paper analyzes  individually-rational ex post equilibrium in
the VC (Vickrey-Clarke) combinatorial auctions. If $\Sigma$ is a
family of bundles of goods, the organizer may restrict the
participants by requiring them to submit their bids only for
bundles in $\Sigma$. The $\Sigma$-VC combinatorial auctions
(multi-good auctions) obtained in this way are known to be
individually-rational truth-telling mechanisms. In contrast, this
paper deals with non-restricted VC  auctions, in which the
buyers restrict themselves to bids on bundles in $\Sigma$, because
it is rational for them to do so. That is,  it may be that when
the buyers report their valuation of the bundles in $\Sigma$, they
are in an equilibrium. We fully characterize those $\Sigma$ that
induce individually rational equilibrium in every VC auction, and
we refer to the associated equilibrium as a bundling equilibrium.
The number of bundles in $\Sigma$ represents the communication
complexity of the equilibrium. A special case of bundling
equilibrium is partition-based equilibrium, in which $\Sigma$ is a
field, that is, it is generated by a partition. We analyze
 the tradeoff between communication complexity and economic
 efficiency of bundling equilibrium, focusing in particular on
 partition-based equilibrium.

\section{Introduction}

The Vickrey-Clarke-Groves (VCG) mechanisms
\cite{Vickrey,Clarke,Groves} are central to the design of
protocols with selfish participants (e.g.,
\cite{NisanRonenGames,Tijcai,Varian95}), and in particular for
combinatorial auctions (e.g.,
\cite{Weber,KrishnaPerry,deVriesVohra,Wellmangeb,MonTenAsymp,Nisan00,Lehmann99}), in which the
participants submit bids, through which they can express
preferences over bundles of goods. The organizer allocates the
goods and collects payments based on the participants'
bids.\footnote{Motivated by the FCC auctions (see e.g.,
\cite{Cramton,McMillan,Milgromfcc} ) there is an extensive recent
literature devoted to the design and analysis of multistage
combinatorial auctions, in which the bidders express partial
preferences over bundles at each stage. See
e.g.,\cite{Wellmangeb,perryrenyb,Ausubel2000,Parkes99,ParkesUngar,AusubelMilgrom}
.} These protocols allow to allocate a set of goods (or services,
or tasks) in a socially optimal (surplus maximizing) manner,
assuming there are no resource bounds on the agents' computational
capabilities.\footnote{There are  at least two sources of
computational issues, which arise when dealing with combinatorial
auctions; Winner determination --finding the optimal allocation
(see e.g.,\cite{RothPekHar,Nisan00,Tenn,Sandholm99,FujiBrownShoham,Anderson,Sandholm01,HoosBoutilier})  , and bid communication --
the transfer of information (see e.g.,
\cite{nisancommunication}).} The VCG protocols are designed in a
way that truth-revealing of the agents' private
information\footnote{This paper deals with the private-values
model, in which every buyer knows his own valuations of bundles
of goods. In contrast, in a correlated-values model, every buyer
receives a signal (possibly about all buyers' valuation
functions), and this signal does not completely reveal his own
valuation function (see e.g.
\cite{MilgromWeber,jehielmsignals,mcafeereny,dasmaskin,perryrenya,perryrenyb}
for
  discussions of
 models
in which valuations are correlated).} is a dominant strategy to
them. Moreover, VCG protocols can be applied in the context of
games in informational form, where no probabilistic assumptions
about agents' types are required.\footnote{ A game in
informational form is a pre-Bayesian game. That is, it has all
the ingredients of a Bayesian game except for the specification
of probabilities. Unlike Bayesian games, games in informational
form do not necessarily possess a solution: a recommendation for
rational players how to play. However, in many important models
such solutions do exist. See Section 2 for a precise definition.}
We shortly define domination and equilibrium in such games. These
solutions are called ex post solutions because they have the
property that if the players were told about the true state,
after they choose their actions, they would not regret their
actions.\footnote{Alternatively, ex post solutions may be called
probability-independent solutions because, up to some
technicalities concerning the concept of measurable sets, they
form Bayesian solutions for every specification of probabilities.}

In this paper we deal with a special type of VCG  mechanisms --
the VC mechanisms. Amongst the VCG mechanisms the VC mechanisms
are characterized by two additional important properties: Truth
telling satisfies the participation constraint, that is, it is
preferred to non participation,\footnote{ An equilibrium that
satisfies the participation constraint  is said to be
individually- rational.} and the seller's revenue is always non
negative.

 A famous observation of the theory of
mechanism design in economics, termed {\bf the revelation
principle} (see e.g. \cite{Myerson79}), implies that the
discussion of additional individually rational (IR) equilibria of
the VC mechanisms may seem unneeded, and indeed it has been
ignored by the literature.  It can be proved that every mechanism
with an ex post equilibrium is economically equivalent to another
mechanism -- a direct mechanism -- in which every agent is
required to submit his information. In this direct mechanism,
revealing the true type is an ex post dominating strategy for
every agent, and it yields the same economics parameters as the
original mechanism. However, the two mechanisms differ in the set
of inputs that the player submits in equilibrium. This difference
may be crucial when we deal with communication complexity. Thus,
two mechanisms that are equivalent from the economics point of
view, may be considered different mechanisms from the CS point of
view.

Thus,  tackling the VC mechanisms from a computational perspective
introduces a vastly different picture. While the revelation of the
agents' types  defines one IR equilibrium, there are other (in
fact, over-exponentially many) IR equilibria for the VC auctions.
Moreover, these equilibria have different communication
requirements.

In this paper we analyze ex post equilibria in the VC mechanisms.

Let $\Sigma$ be a family of bundles of goods. We characterize
those $\Sigma$, for which the strategy of reporting the true
valuation over the bundles in $\Sigma$ is a player-symmetric IR ex
post equilibrium.  An equilibrium that is defined by such $\Sigma$
is called a bundling equilibrium. We prove that $\Sigma$ induces a
bundling equilibrium if and only if  it is a quasi field of
bundles.\footnote{A quasi field is a nonempty set of sets that is
closed under complements and under disjoint unions.} The number of
bundles in $\Sigma$ represents the communication complexity of the
equilibrium, and the  economic efficiency of an equilibrium is
measured by the generated social surplus.
   A special type of bundling
equilibria are partition-based equilibria, in which $\Sigma$ is a
field (i.e. it is generated by a partition).  The partition-based
equilibria are ranked according to the usual partial order on
partitions: If one partition is finer than another one, then it
yields higher communication complexity as well as higher social
surplus.\footnote{  It is worth mentioning that the various
equilibria cannot be ranked according to the revenue of the
seller. That is, under some conditions, a partition-based
equilibrium may simultaneously yield more revenue and less
communication complexity than the truth revealing equilibrium (an
example is provided).}

We analyze the least upper bound  (over all possible profiles of
valuation functions, one for each buyer) of the ratio between the
optimal surplus and the surplus obtained in a partition-based
equilibrium. We express this least upper bound in terms of the
partition's structure. We provide an upper bound for this ratio,
which is proved to be tight in infinitely many cases.

In Section~2 we present the concept of ex post equilibrium in
games in informational form. In Section 3 we discuss combinatorial
auctions. Together, Sections 2 and 3 provide the reader with a
rigorous framework for general analysis of VC protocols for
combinatorial auctions.  In Section 4  we introduce bundling
equilibrium, and provide a full characterization of bundling
equilibria for VC protocols.
 Then we discuss bundling equilibrium that is generated by a partition,
 titled  partition-based equilibrium.
 In Section 5 we deal with the surplus of VC protocols
for combinatorial auctions
when following partition-based equilibrium, exploring the spectrum between
economic efficiency and communication efficiency.

\section{Ex post equilibrium in games in informational form}
 A game in  {\bf informational form} $G=G(N,\Omega,T,(\tilde t_i)_{i\in N},X,(u_i)_{i\in N})$
  is defined
 by the following parameters:

 $\bullet$ {\bf Agents}: Let $N=\{1,\ldots,n\}$ be the
 set of agents.

$\bullet$ {\bf States}: Let $\Omega$ be the set of  (relevant)
states.

$\bullet$ {\bf Types}: Let $T_i$ be the set of types of agent $i$,
$T=\times_{i\in N}T_i$.

$\bullet$ {\bf Signaling functions}: Let $\tilde t_i:\Omega\to
T_i$ be the signaling function of agent $i$. Without loss of
generality, it is assumed that every type $t_i\in T_i$ is
possible. That is $\tilde t_i(\Omega)=T_i$.

$\bullet$ {\bf Actions}: Let $X_i$ be the set of actions of $i$,
$X=\times_{i\in N}X_i.$

$\bullet$ {\bf Utility functions}: Let $u_i(\omega, x)$ be the
utility of $i$ at state $\omega$, when the agents choose the
action profile $x$.

 Every $\omega\in\Omega$ defines a game in strategic
(normal) form, $G(\omega)$. In this game agent $i$ receives
$u_i(\omega, x)$, when it chooses $x_i$, and all other agents
choose $x_{-i}$. However, the agents do not know which game they
play.

For $t_i\in T_i$ let $\Omega_i(t_i)$ be the set of states that
generate the signal $t_i$, that is
 $$\Omega_i(t_i)=\{\omega\in\Omega|\tilde t_i(\omega)=t_i\}.$$

  A {\bf strategy}\footnote{In this paper
  we do not deal with mixed strategies.}
   of $i$ is a function $b_i:T_i\to X_i$;
 The associated {\bf implied strategy} is the function $\hat
 b_i:\Omega\to X_i$ given by
 $$\hat b_i(\omega)=b_i(\tilde t_i(\omega)).$$

 A profile of strategies $b=(b_1,\cdots,b_n)$ is an {\bf
 ex post equilibrium}, if for every agent $i$, for every $t_i\in T_i$, for
every $\omega\in\Omega_i(t_i)$, and for every $x_i\in X_i$,
$$u_i(\omega,b_i(t_i),\hat b_{-i}(\omega))\ge u_i(\omega,x_i,\hat
b_{-i}(\omega)).$$

A strategy $b_i$ of $i$ is an {\bf ex post dominant strategy} for
$i$, if for every profile of strategies $b_{-i}$ of the other
players, for every  $t_i\in T_i$, for every
$\omega\in\Omega_i(t_i)$, and for every $x_i\in X_i$,
$$u_i(\omega,b_i(t_i),\hat b_{-i}(\omega))\ge u_i(\omega,x_i,\hat
b_{-i}(\omega)).$$ Obviously, if $b_i$ is an  ex post dominant
strategy for every $i$, $b$ is an ex post equilibrium, but not
necessarily  vice versa. An ex post equilibrium $b$, in which
every strategy $b_i$ is ex post dominant is called an ex post {\bf
domination equilibrium}.

\section{Combinatorial auctions}
In a combinatorial auction there is a seller, denoted by $0$, who
wishes to sell a set of  $m$ items $A=\{a_1,\ldots,a_m\}$, $m\ge
1$, that are owned by her. There is a set of (potential) buyers
$N=\{1,2,\ldots,n\}$, $n\ge 1$. We take $N$ as the set of agents.
Let $\Gamma$ be the set of all allocations of the goods. That is,
every $\gamma\in \Gamma$ is an ordered partition of $A$,
$\gamma=(\gamma_i)_{i\in N\cup\{0\}}$.
A {\bf valuation function} of buyer $i$ is a function $v_i:2^A\to
\Re$, where $\Re$ denotes the set of real numbers, with the
normalization $v_i(\emptyset)=0$. In a more general setup, a buyer
may care about the distribution of goods that he does not own. In
such a setup the utility  of an agent may depend on the whole
allocation $\gamma$ rather than on $\gamma_i$. Hence, by dealing
with valuation functions we actually assume:

$\bullet$ {\bf No allocative externalities}.\footnote{For auctions
in which externalities are assumed see, e.g.,
\cite{jmols,jmolsAER}.}

We also assume:

$\bullet$ {\bf Free disposal}: If $B\subseteq C$, $B,C\in 2^A$,
then $v_i(B)\le v_i(C)$.

Let $V_i$ be the set of all possible valuation functions of $i$
(obviously $V_i=V_j$ for all $i,j\in N$), and let $V=\times_{i\in
N}V_i$. We refer to $V$ as the set of {\bf states} ($\Omega=V$).
In a general model, each buyer receives a signal $t_i$  through a
signaling function $\tilde t_i$ defined on $V$. We assume:

$\bullet$  {\bf Private value model}: $\tilde t_i(v)=v_i$. That
is, $T_i=V_i$ and each buyer knows his valuation function
only.\footnote{ See e.g.
 \cite{MilgromWeber,jehielmsignals,mcafeereny,dasmaskin,perryrenya,perryrenyb} for
  discussions of
 models
in which valuations are correlated and buyers do not know their
own valuation.}

A {\bf mechanism}  for allocating the goods is defined by sets of
messages $X_i$, one set for each buyer $i$, and by a pair $(d,c)$
with $d:X\to \Gamma$, and $c:X\to \Re^n$, where $X=\times X_i$.
$d$ is called the allocation function and $c$ the transfer
function; if the buyers send the profile of messages $x\in X$,
buyer $i$ receives the set of goods $d_i(x)$ and pays $c_i(x)$ to
the seller. We assume:

$\bullet$  {\bf Quasi linear utilities}: If agent $i$ with the
valuation function $v_i$ receives the set of goods $\gamma_i$ and
pays $c_i$, his utility equals $v_i(\gamma_i)-c_i$.

As the seller cannot force the buyer to participate, a full
description of a mechanism should describe the allocation of goods
and transfers for cases in which not all agents participate.
However, we adopt the way this issue is treated in economics: The
mechanism $(X,d,c)$ defines a game in informational form. An ex
post equilibrium $b$ in this game satisfies the {\bf participation
constraint} if for every buyer $i$, $$v_i(d_i(b(v)))-c_i(b(v))\ge
0\quad\mbox{for every $v \in V$,}\eqno{(3.1)}$$ where
$b(v)=(b_1(v_1), \ldots ,b_n(v_n))$. If an ex post equilibrium
satisfies the participation constraint, we call it {\bf
individually rational}. If the buyers use an individually rational
ex post equilibrium profile $b$, then a deviation of a buyer to
non participation is not profitable for him.\footnote{ Thus, $b$
remains an ex post equilibrium profile if every set of messages is
extended by a null message, and an agent whose input is null
receives no good and pays nothing. Nevertheless, if the issue of
uniqueness of equilibrium is important, dealing with ex post
individually rational equilibrium instead of dealing with ex post
equilibrium in the extended model is not without loss of
generality. The extended model may have more equilibrium profiles,
that cannot be expressed in the reduced model, that is,
equilibrium profiles in which some of the agents do not
participate in some of the cases.} Similarly, a dominant strategy
of $i$, $b_i$, is individually rational if (3.1) is satisfied for
every profile $b_{-i}$ of the other buyers' strategies.

 For an allocation $\gamma$ and a profile of types
$v$ we denote by $S(v,\gamma)$ the {\bf total social surplus} of
the buyers, that is $$S(v,\gamma)=\sum_{i\in N} v_i(\gamma_i).$$
We also denote: $$S_{max}(v)=\max_{\gamma\in \Gamma}S(v,\gamma).$$

Consider a mechanism $M=(X,d,c)$ and an individually rational ex
post equilibrium $b$.
 For every profile $v$ we denote the surplus generated by $b$ by
$S^M_b(v)=S(v,d(b(v)))$, and the revenue collected by the seller
by $R^M_b(v)=\sum_{i\in N}c_i(b(v))$.

Because of the participation constraint, $$R^M_b(v)\le S^M_b(v)\le
S_{max}(v)\quad\mbox{for all $v\in V$.}$$ A mechanism and an
individually rational ex post equilibrium $(M,b)$ are called {\bf
socially optimal} if $$S^M_b(v)=S_{max}(v)\quad\mbox{ for all
$v\in V$.}$$

Note that the seller controls the mechanism, but she does not
control the strategies used by the buyers. However, it is assumed
that if the mechanism possesses an individually rational ex post
equilibrium, the agents use such an equilibrium.\footnote{ For
example,  the agents may reach the equilibrium by a process of
learning  (see e.g. \cite{HonSela}).}

A public seller may wish to generate a socially optimal mechanism,
whereas a selfish seller may be interested in the revenue function
only. Such a seller would rank mechanisms according to the revenue
they generate.

A mechanism $(X,d,c)$ is called a  {\bf direct} mechanism if
$X_i=V_i$ for every $i\in N$. That is, in a direct mechanism a
buyer's message contains a full description of some valuation
function. A direct mechanism is called {\bf truth revealing} if
for every buyer $i$, telling the truth ($b_i(v_i)=v_i$) is an
 individually rational ex post dominating strategy. (Of course the profile of
 strategies $b=(b_i)_{i\in N}$ is an individually rational
ex post equilibrium.) By the revelation principle,\footnote{See
e.g. \cite{ Myerson79}.} given a mechanism and an individually
rational ex post equilibrium $(M,b)$ one can find a direct truth
revealing mechanism that yields the same distribution of goods and
the same payments (and in particular, the same revenue and surplus
functions).\footnote{ This strong version of the revelation
principle is due to our private values assumption. Otherwise, the
revelation principle guarantees the existence of an equivalent
direct mechanism in which telling the truth is an individually
rational ex post equilibrium (but not necessarily a domination
equilibrium).} It may seem therefore that the concept of ex post
equilibrium is not interesting in our setup (private values), and
indeed most of the economics literature of mechanism design with
private values deals only with direct and truth revealing
mechanisms.
 However, when we deal with computational issues, two mechanism
that are equivalent in economics may differ in their complexity.
The time, space, and  communication required to compute and communicate
the message of
an agent  as well as the chosen allocation may depend on the
messages sent in equilibrium. Thus, the concept of ex post
equilibrium may be very important even if private values are
assumed.

 Well-known  truth revealing mechanisms are the VC
mechanisms.  These mechanisms are parameterized by an allocation
function $d$, that is socially optimal. That is,
$S_{max}(v)=S(v,d(v))$ for every $v\in V$.  The transfer functions
are defined as follows: $$c^d_i(v)=\max_{\gamma\in
\Gamma}\sum_{j\ne i}v_j(\gamma_j)-\sum_{j\ne
i}v_j(d_j(v)).\eqno{(3.2)}$$ Note that $c_i^d(v) \ge 0$ for every
$v \in V$.

The mechanisms differ in the allocation they pick in cases in
which there exist more than one socially optimal allocation, and
therefore in the second term in (3.2).\footnote{By VC mechanisms
we refer here to what is also known as Clarke mechanisms or the
Pivotal mechanism. More general mechanisms are the VCG mechanisms.
Every VCG mechanism is obtained from some VC mechanism by changing
the transfer functions: A VCG mechanism is defined by a socially
optimal allocation function $d$ and by a family of functions
$h=(h_i)_{i\in N}$. The transfer functions are defined by:
$$c^d_i(v)=\max_{\gamma\in \Gamma}\sum_{j\ne
i}v_j(\gamma_j)-\sum_{j\ne i}v_j(d_j(v))+h_i(v_{-i}).$$ Truth
telling is an ex post equilibrium in every VCG mechanism, but it
is not necessarily an individually rational ex post equilibrium.}
It is well-known that {\bf all VC mechanisms yield the same
utility to a truth telling buyer}: For a VC mechanism
$d$\footnote{Since in all VC mechanisms $M=(X,d,c)$, $X$ is $V$
and $c$ is defined as in (3.2), it is enough to specify $d$ in
order to specify the mechanism.} we denote by
$u_i^d(v_i,(v'_i,v_{-i}))$ the utility of buyer $i$ with the
valuation function $v_i$, when he declares $v'_i$ and the other
buyers declare $v_{-i}$. That is,
$$u_i^d(v_i,v')=v_i(d_i(v'))-c^{d}_i(v'),\eqno{(3.3)}$$
 where $v'=(v'_i,v_{-i})$. Therefore, by (3.2),
$$u_i^d(v_i,v')=S(v,d(v'))-g_i(v_{-i}),\eqno{(3.4)}$$
where $v=(v_i,v_{-i})$, and
$$g_i(v_{-i})=\max_{\gamma\in \Gamma}\sum_{j\ne
i}v_j(\gamma_j).\eqno{(3.5)}$$
 If $i$ declares $v_i$,
$$ u_i^d(v_i,v)=S(v,d(v))-g_i(v_{-i})=S_{max}(v)-g_i(v_{-i}).\eqno{(3.6)}$$
As the right-hand side of (3.6) does not depend on $d$, a truth
telling buyer receives the same utility at all VC mechanisms.
Note that truth revealing is indeed a dominant strategy in a VC
mechanism.

In the next section we discuss other (not truth revealing)
individually rational ex post equilibrium profiles in the VC
mechanisms. We will focus on {\bf player symmetric} equilibria
$b=(b_i)_{i \in N}$, where $b_i=b_j$ for all $i,j\in N$, which are
in equilibrium in {\bf every} VC mechanism.

\section{Bundling equilibrium}
Let $\Sigma\subseteq 2^A$ be a family of bundles of goods. We
deal only with such families $\Sigma$ for which
\begin{itemize}
\item $\emptyset\in\Sigma$.
\end{itemize}
A valuation function $v_i$ is a {\bf $\Sigma$-valuation function}
if $$v_i(B)=\max_{C\in\Sigma,C\subseteq B}v_i(C),\quad\mbox{for
every $B\in 2^A$.}$$

The set of all $\Sigma$-valuation functions in $V_i$ is denoted by
$V_i^\Sigma$. We further denote $V^\Sigma=\times_{i\in
N}V_i^\Sigma$. For every valuation function $v_i$ we denote by
$v_i^\Sigma$ its projection on $V_i^\Sigma$, that is:
$$v_i^\Sigma(B)=\max_{C\in\Sigma,C\subseteq
B}v_i(C),\quad\mbox{for every $B\in 2^A$.}$$ Obviously
$v_i^\Sigma\in V_i^\Sigma$, and for $v_i\in V_i^\Sigma$,
$v_i^\Sigma=v_i$. In particular $(v_i^\Sigma)^\Sigma=v_i^\Sigma$
for every $v_i\in V_i$. Let $f_\Sigma:V_i\to V_i^\Sigma$ be the
projection function defined by $$f^\Sigma(v_i)=v_i^\Sigma.$$

 An
allocation $\gamma$ is a {\bf $\Sigma$-allocation} if
$\gamma_i\in\Sigma$ for every buyer $i\in N$. The set of all
$\Sigma$-allocations is denoted by $\Gamma^\Sigma$.

We are interested in the following question: For which $\Sigma$,
do we have that $f^\Sigma$ is a player-symmetric individually
rational ex post equilibrium in every   VC mechanism (with any
number of buyers)? In such a case we call $f^\Sigma$ a {\bf
bundling equilibrium }for the VC mechanisms and say that $\Sigma$
induces a bundling equilibrium. The next example shows that not
every $\Sigma$ induces a bundling equilibrium.

Before we present the example we need the following notation: Let
$B\in 2^A$, we denote by $w_B$ the following valuation function:
$$\quad\mbox{If $B\neq \emptyset$, } w_B(C)=1\quad\mbox{if
$B\subseteq C$, and}\quad w_B(C)=0
\quad\mbox{otherwise}.\footnote{For $B\ne\emptyset$, a valuation
function of the form $w_B$ is called a unanimity TU game in
cooperative game theory. An agent with such a valuation function
is called by Lehmann, O'Callaghan, and Shoham \cite{Lehmann99} a
single-minded agent.}$$ $$\quad\mbox{If $B=\emptyset$, } w_B(C)=0
\quad\mbox{for all } C \in 2^A.$$

\noindent{\bf Example 1}

Let $A$ contain four goods $a,b,c,d.$ Let $$\Sigma=\{a,d, bcd,
abc, A,\emptyset\}.\footnote{We omit braces and commas when
writing subsets of $A$.}$$ Let $v_2=w_{a}, v_3=w_{d}$. Consider
buyer 1 with $v_1=w_{bc}$. Note that $v_i\in V_i^\Sigma$ for
$i=2,3$. If buyer 1 uses $f^\Sigma$ he declares $
v_1'(bcd)=v_1'(abc)=v_1'(A)=1$ and $v_1'(C)=0$ for all other $C$,
 and there
exists a VC mechanism  that allocates $a$ to 2, $d$ to 3 and $bc$
to the seller. In this mechanism the utility of 1 from using
$f^\Sigma$ is zero. On the other hand, if agent 1 reports the
truth ($w_{bc}$) he receives  (in every VC mechanism) $bc$ and
pays nothing. Hence, his utility would be 1. Therefore $f^\Sigma$
is not in equilibrium in this VC mechanism, and hence $\Sigma$
does not induce a bundling equilibrium.

\subsection{A characterization of bundling equilibria}
$\Sigma\subseteq 2^A$ is called a {\bf quasi field} if it
satisfies the following properties:\footnote{Recall our assumption
that we deal only with $\Sigma$ such that $\emptyset\in\Sigma$.}
\begin{itemize}
\item $B\in \Sigma$ implies that $B^c\in \Sigma$, where
$B^c=A\setminus B$.
\item $B,C \in \Sigma$ and $B \cap C= \emptyset$ imply that $B
\cup C \in \Sigma$.\footnote{Equivalently, the union of any number
of pairwise disjoint sets in $\Sigma$ is also in $\Sigma$.}
\end{itemize}

\begin{theorem} $\Sigma$ induces a bundling equilibrium if and only if
it is a quasi field.
\end{theorem}

\begin{proof}

{\bf Suppose $\Sigma$ is a quasi field}:

 Consider a VC mechanism with
an allocation function $d$. We show that $f^\Sigma$ is an
individually rational ex post equilibrium in this VC mechanism.

Assume that every buyer $j$, $j\ne i$, uses the strategy
$b_j=f^\Sigma$. Let $v_{-i}\in V_{-i}$. We have to show that for
buyer $i$ with valuation $v_i$,  $v_i^\Sigma$ is a best reply to
$v_{-i}^\Sigma$. As truth revealing is a dominating strategy in
every VC mechanism, it suffices to show that buyer $i$'s utility
when submitting $v_i^\Sigma$ is the same as when submitting
$v_i$.\footnote{Note that this will imply not only that $f^\Sigma$
is in equilibrium but also that it is individually rational.} That
is, we need to show that

$$S_{max}(v_i,v^\Sigma_{-i})-\alpha=S((v_i,v^\Sigma_{-i}),\gamma)-\alpha,$$
where $\alpha=g_i( v_{-i}^\Sigma)$, and
$\gamma=d(v_i^\Sigma,v^\Sigma_{-i})$.

Hence, we have to show that
$$S_{max}(v_i,v^\Sigma_{-i})=S((v_i,v^\Sigma_{-i}),\gamma).\eqno{(4.1)}$$
Obviously,
$$S_{max}(v_i,v^\Sigma_{-i})\ge S((v_i,v^\Sigma_{-i}),\gamma).\eqno{(4.2)}$$

 As $v_i(B)\ge v_i^\Sigma(B)$ for every $B\in 2^A$,
$$S((v_i,v^\Sigma_{-i}),\gamma)\ge
S((v_i^\Sigma,v^\Sigma_{-i}),\gamma)=S_{max}(v_i^\Sigma,v^\Sigma_{-i}).\eqno{(4.3)}$$

Let $ \xi=d(v_i,v^\Sigma_{-i})$. For $j\ne i$ and $j\ne 0$, let
$\xi_j^\Sigma\in\Sigma$ be such that $\xi^\Sigma_j\subseteq \xi_j$
and $v^\Sigma_j(\xi^\Sigma_j)=v^\Sigma_j(\xi_j).$ Let
$\xi^\Sigma_i=(\cup_{j\ne 0,i} \xi^\Sigma_j)^c$, and let
$\xi_0^\Sigma=\emptyset$.

 Because $\Sigma$ is a quasi field,
$\xi^\Sigma_i\in\Sigma$, and hence $\xi^\Sigma\in\Gamma^\Sigma$.
As $\xi_i\subseteq \xi^\Sigma_i$, $\xi^\Sigma$ is also optimal for
$(v_i,v^\Sigma_{-i})$. However
$$S((v_i,v^\Sigma_{-i}),\xi^\Sigma)=S((v_i^\Sigma,v^\Sigma_{-i}),\xi^\Sigma)\le
S_{max}(v_i^\Sigma,v^\Sigma_{-i}).\eqno{(4.4)}$$ Combining (4.2),
(4.3), and (4.4) yields
$$S_{max}(v_i^\Sigma,v^\Sigma_{-i}) \ge S_{max}(v_i,v^\Sigma_{-i}) \ge
S((v_i,v^\Sigma_{-i}),\gamma) \ge
S_{max}(v_i^\Sigma,v^\Sigma_{-i}).$$ Therefore (4.1) holds.

{\bf Suppose $\Sigma$ induces a bundling equilibrium}:

We first show that if $B\in\Sigma$, then $B^c\in\Sigma$.  If $B=A$
then by definition $B^c=\emptyset \in \Sigma$. Let $B\subset A$.
Assume, for the sake of contradiction, that $B^c\not\in\Sigma$.
Let $v_2=w_B$ and $v_1=w_{B^c}$. Note that $v_2^\Sigma=v_2$. Thus,
if buyer 2 uses $f^\Sigma$, he declares $v_2$. If buyer 1 uses
$f^\Sigma$, he declares $v_1^\Sigma$, where $v_1^\Sigma(B^c)=0$.
Hence, there exists a VC mechanism $d$, that allocates $B$ to
agent 2 and $B^c$ to the seller. However, if buyer 1 deviates and
declares his true valuation, then this VC mechanism allocates to
him $B^c$, and he pays nothing. Hence, there is a profitable
deviation from $f^\Sigma$, a contradiction.

Next, we show that if $B,C \in \Sigma$ are disjoint then $B \cup C
\in \Sigma$. By the first part of the proof, it suffices to show
that $(B \cup C)^c \in \Sigma$. Clearly, we may assume that the
sets $B$, $C$, and $(B \cup C)^c$ are all non empty. Assume, for
the sake of contradiction, that $(B \cup C)^c \notin \Sigma$.
Consider three buyers with valuations $v_1=w_{(B \cup C)^c}$,
$v_2=w_B$, $v_3=w_C$. Proceeding as in the
 first part of the current part of the  proof yields a similar
 contradiction.\QED
 \end{proof}

 It may be useful to note that if $f^\Sigma$ is a buyer-symmetric
 equilibrium for a fixed set of buyers, then $\Sigma$ is not
 necessarily a quasi field. For example, if there is only one
 buyer, every $\Sigma$ such that $A\in\Sigma$ induces an
 equilibrium. In the case of two buyers, being closed under
 complements is necessary and sufficient for $\Sigma$ to induce
 an equilibrium. However, it can be deduced from the proof of the
 only if part of Theorem 1, that for a fixed set of buyers $N$,
 if $n=|N| \ge 3$, then $\Sigma$ must be a quasi field if it induces
 an equilibrium for the set of buyers $N$.

\subsection{Partition-based equilibrium}
Let $\pi=\{A_1,...,A_k\}$ be a partition of $A$ into non empty
parts. That is, $A_i\ne\emptyset$ for every $A_i\in\pi$,
$\cup_{i=1}^k A_i=A$, and $A_i\cap A_j=\emptyset$ for every $i\ne
j$. Let $\Sigma_\pi$ be the field generated by $\pi$. That is,
$\Sigma_\pi$ contains all the sets of goods of the form
$\cup_{i\in I} A_i$, where $I\subseteq\{1,...,k\}$. To avoid
confusion: $\emptyset\in \Sigma_\pi$. For convenience, we will
 use $f^\pi$ to denote
$f^{\Sigma_\pi}$.  A
corollary of Theorem 1 is:
\begin{corollary}
$f^\pi$ is a bundling equilibrium.
\end{corollary}
\begin{proof}
As $\Sigma_\pi$ is a field it is in particular a quasi field.
Hence, the proof follows from Theorem 1.\QED
\end{proof}

A bundling equilibrium of the form $f^\pi$, where $\pi$ is a
partition, will be called a {\bf partition-based equilibrium}.
Thus, a partition-based equilibrium is a bundling equilibrium
$f^\Sigma$ that is based on a field $\Sigma=\Sigma_\pi$.
 It is important to note that there exist quasi fields, which are
 not fields.
 For example, let $A=\{a,b,c,d\}$. $\Sigma=\{ab, cd, ac, bd, A ,
\emptyset\}$ is a quasi field, which is not a field. We note,
however, that when $m=|A| \le 3$, the notions of quasi field and
field coincide.

\section{Surplus and communication complexity}
 Let $\Sigma\subseteq 2^A$. If every buyer uses
$f^\Sigma$, then in every VC mechanism $d$, the total surplus
generated when the types of the buyers are given by $v\in V$ is
$S(v^\Sigma,d(v^\Sigma))=S_{max}(v^\Sigma)$.

 We denote
$$S_{\Sigma-max}(v)=\max_{\gamma\in\Gamma^\Sigma}S(v,\gamma).$$
Obviously,

$$S_{\Sigma-max}(v)=S_{\Sigma-max}(v^\Sigma)=S_{max}(v^\Sigma),\quad\mbox{for
every $v\in V$.}$$



For convenience we denote $S_{\Sigma-max}$ by $S_\Sigma$, and we
call $S_\Sigma$ the {\bf $\Sigma$-optimal surplus function} (note
that $S_{2^A}=S_{max}$). When $\Sigma$ is a field generated by a
partition $\pi$ we write $S_\pi$ for $S_{\Sigma_\pi}$.

If $\Sigma$ is a quasi field we say that the {\bf communication
complexity} of the equilibrium $f^\Sigma$ is the number of bundles
in $\Sigma$, that is $|\Sigma|$. Notice that this is a natural definition
because a buyer who is using $f^\Sigma$ has to submit  a vector of
$|\Sigma|$ numbers to the seller.%
\footnote{A discussion of the way this can be extended to deal with
the introduction of concise bidding languages \cite{Nisan00,BoutilierHoos}
 is beyond the scope of this
paper.} Thus, if $\pi$ is a partition, the communication
complexity is $2^{|\pi|}$. If $\Sigma_1\subseteq\Sigma_2$, then
$S_{\Sigma_1}(v)\le S_{\Sigma_2}(v)$ for every $v\in V$. So,
$\Sigma_2$ induces more surplus (a proxy for economic efficiency)
than $\Sigma_1$, but $\Sigma_2$ also induces higher  communication
complexity.
Hence, there is a tradeoff between
  economic efficiency and computational
 complexity.
 The next example shows that as far as the {\bf revenue} of the
 seller is concerned,
  there is no clear
 comparison between the revenues obtained by quasi
 fields ranked by inclusion.\footnote{ In spite of our example, it is commonly
 believed that social optimality is a good proxy for revenue.
 This was proved to be asymptotically correct when the number of
 buyers is large, and the organizer has a Bayesian belief over the
 distribution of valuation functions, which assumes independence
 across buyers (see \cite{MonTenAsymp}).}
 Before we present the example, note that for two partitions
 $\pi_1,\pi_2$, $\Sigma_{\pi_1}\subseteq\Sigma_{\pi_2}$ if and
 only if $\pi_2$ refines $\pi_1$.

 \noindent{\bf Example 2}

 Assume there are two buyers, $N=\{1,2\}$, and two goods, $A=\{a,b\}$.
Assume $v_1=w_a$ and $v_2=w_b$. In any VC mechanism, in the truth
revealing equilibrium buyer 1 gets $a$, buyer 2 gets $b$, and they
pay nothing. Hence the level of social surplus is $2$ and the
revenue of the seller at $v=(v_1,v_2)$ is zero.
 Let $\pi$ be the trivial partition $\{A\}$
 ($\Sigma_\pi=\{\emptyset,A\}$).
If  each  buyer uses  the equilibrium strategy $f^{\pi}$, they
both report $w_A$. Hence, one  of the buyers gets $ab$ and pays 1.
The seller collects a revenue of $1$, and the social surplus
equals 1. Hence, $S_{max}(v)>S_\pi(v)$ and $R(v)<R_\pi(v)$. On the
other hand, if $N=\{1,2,3,4\}$ where $v_1$ and $v_2$ are defined
as before and $v_3=v_1$ and $v_4=v_2$, $S_{max}(v)=2$ and $R(v)=2$
while $S_\pi(v)=1$ and $R_\pi(v)=1$.

For every family of bundles $\Sigma$ with $A\in\Sigma$ we define
$$r_\Sigma^n=\sup_{v\in V,v\ne 0}\frac{
S_{max}(v)}{S_{\Sigma}(v)},\eqno{(5.1)}$$ where $V=V_1 \times
\cdots \times V_n$. Thus, $r_\Sigma^n$ is a worst-case measure of
the economic inefficiency that may result from using the strategy
$f^\Sigma$ when there are $n$ buyers. Obviously $r_\Sigma^n \ge
1$, and equality holds for $\Sigma = 2^A$. A standard argument
using homogeneity and continuity of $S_{max}/S_\Sigma$ shows that
the supremum in (5.1) is attained, i.e., it is a maximum.

The following remark gives a simple upper bound on the
inefficiency associated with $\Sigma$.

\noindent {\bf Remark 1}
 For every $\Sigma \subseteq 2^A$ with $A \in \Sigma$, and for every $v\in V$,
 $$S_{max}(v)\le n S_\Sigma(v),$$
 where $n$ is the number of potential buyers.
Consequently,
 $$r^n_\Sigma\le n.$$

 \begin{proof}
 Let $\gamma=d(v)$, where $d$ is any VC mechanism.
 $$S_{max}(v)=S(v,\gamma)=\sum_{i\in N}v_i(\gamma_i)\le \sum_{i\in N}v_i(A)=
 \sum_{i\in N}v^\Sigma_i(A)
 \le n S_\Sigma(v).$$\QED
 \end{proof}

However, we are interested mainly in upper bounds on the economic
inefficiency that are independent of the number of buyers. For
every family of bundles $\Sigma$ with $A \in \Sigma$ we define $$
r_\Sigma = \sup_{n \ge 1} r_\Sigma^n.\eqno{(5.2)}$$ It is easy to
see that, since any allocation assigns non empty bundles to at
most $m=|A|$ buyers, the supremum in (5.2) is attained for some $n
\le m$. When $\Sigma = \Sigma_\pi$ for a partition $\pi$, we write
$r_\pi$ instead of $r_{\Sigma_\pi}$.

In the following subsection we characterize and estimate $r_\pi$,
thereby obtaining a quantitative form of the tradeoff between
communication and economic efficiency in partition-based
equilibria.

 \subsection{ Communication efficiency vs.
 economic efficiency in partition-based equilibria}

 We first express $r_\pi$ in
terms of the partition $\pi = \{A_1,...,A_k\}$ only. A {\bf
feasible} family for $\pi$ is a family $\Delta =(H_i)_{i=1}^s$ of
(not necessarily distinct) subsets of $\{1,...,k\}$ satisfying the
following two conditions:

\begin{itemize}
\item
 $H_i\cap
H_j\ne\emptyset$ for every $1\le i,j\le s$.
\item $|\{i:l \in H_i \}| \le |A_l|$ for every $1 \le l \le k$.
\end{itemize}
We write $s=s(\Delta)$ for the number of sets in the family
$\Delta$ (counted with repetitions).

\begin{theorem}
For every partition $\pi$, $$r_\pi= \max s(\Delta),$$ where the
maximum is taken over all families $\Delta$ that are feasible for
$\pi$.
\end{theorem}
\begin{proof}

We first prove that $r_\pi \le \max s(\Delta)$. It suffices to
show that for every $v \in V$ there exists a feasible family
$\Delta$ for $\pi$ such that $$S_{max}(v) \le s(\Delta) \cdot
S_\pi(v).$$

Let $v\in V$. Let $\gamma$ be a socially optimal allocation. That
is, $$S_{max}(v)=\sum_{i\in N}v_i(\gamma_i).$$ For every
$\gamma_i$ let $\gamma_i^\pi$ be the minimal set in $\Sigma_\pi$
that contains $\gamma_i$. That is $\gamma_i^\pi=\cup_{l\in
J_i}A_l$ where $J_i=\{l \in \{1,\ldots,k\}:A_l \cap \gamma_i \neq
\emptyset\}$.

Let $\xi$ be a partition of $N$ to $r$ subsets, such that for
every $i,j\in I\in\xi$, $i\ne j$, $\gamma_i^\pi\cap
\gamma_j^\pi=\emptyset$. Assume $r$ is the {\bf minimal}
cardinality of such a partition. For every $I\in\xi$ let
$H_I=\cup_{i \in I} J_i$. That is, each $H_I$ is a set of indices
of parts $A_l$ in $\pi$ that should be allocated to the buyers in
$I$ in order for each of them to get the goods they received in
the optimal allocation $\gamma$. Note that if $I\ne J$, $H_I\cap
H_J\ne\emptyset$, otherwise we can join $I$ and $J$ together in
contradiction to the minimality of  the cardinality of $\xi$.
Hence, $\Delta =(H_I)_{I\in\xi}$ is a family of subsets of
$\{1,...,k\}$ that satisfies that any two subsets in $\Delta$
intersect. Furthermore, the second condition for feasibility is
also satisfied, because for any given $l \in \{1,...,k\}$ there
are at most $|A_l|$ buyers $i$ with $\gamma_i \cap A_l \ne
\emptyset$, and hence at most $|A_l|$ parts $I \in \xi$ such that
$l \in H_I$. Thus, $\Delta$ is a feasible family for $\pi$ with
$s(\Delta)=r$.

 Every $H_I$, $I\in\xi$ defines a
$\Sigma_\pi$-allocation. In this allocation every $i\in I$
receives $\gamma_i^\pi$, and the seller receives all other goods.
 Therefore $\sum_{i\in I}v_i(\gamma_i^\pi)\le S_\pi(v)$ for every
 $I\in\xi$. Hence,
 $$S_{max}(v)\le \sum_{i\in
 N}v_i(\gamma_i^\pi)=\sum_{I\in\xi}\sum_{i\in
 I}v_i(\gamma_i^\pi)\le\sum_{I\in\xi}S_\pi(v)=rS_\pi(v).$$

Next, we prove that $r_\pi \ge \max s(\Delta)$. It suffices to
show that for every feasible family $\Delta$ for $\pi$ there
exists a profile of valuations $v=(v_1,...,v_n) \ne 0$ for some
number $n$ of buyers satisfying $$S_{max}(v) \ge s(\Delta) \cdot
S_\pi(v).$$

Let $\Delta =(H_i)_{i=1}^s$ be a feasible family for $\pi$. By the
second condition of feasibility, we can associate with each $H_i$
a set of goods $B_i$ containing one good from each $A_l$ such that
$l \in H_i$, in such a way that the sets $B_i$ are pairwise
disjoint. By the first condition of feasibility, for every $1 \le
i,j \le s$ there can be no two disjoint sets $C_i, C_j \in
\Sigma_\pi$ such that $B_i \subseteq C_i$, $B_j \subseteq C_j$.

Now, we take $n=s$ buyers, and let buyer $i$ have the valuation
$v_i=w_{B_i}$. Then $S_{max}(v)=s$ whereas $S_\pi(v)=1$.\QED
\end{proof}

Theorem~2 reduces the determination of the economic inefficiency
measure $r_\pi$ to a purely combinatorial problem. However, this
combinatorial problem does not admit an easy
solution.\footnote{The special case of this problem, in which
$|A_i|=|A_j|$ for all $A_i, A_j \in \pi$, has been treated in the
combinatorial literature using a different but equivalent
terminology (see e.g. \cite{Furedi}). But even in this case, a
precise formula for $\max s(\Delta)$ seems out of reach.}
Nevertheless, we will use Theorem~2 to calculate $r_\pi$ in some
special cases, and to obtain a general upper bound for it which is
tight in infinitely many cases.

The following proposition determines $r_\pi$ for partitions $\pi$
with a small number of parts. We use the notations $\lfloor \cdot
\rfloor$ and $\lceil \cdot \rceil$ for the lower and upper integer
rounding functions, respectively.

\begin{proposition}
Let $|A|=m$, and let $\pi = \{A_1,...,A_k\}$ be a partition of $A$
into $k$ non empty sets.
\begin{itemize}
\item If $k=1$ then $r_\pi=m$.
\item If $k=2$ then $r_\pi= \max \{|A_1|,|A_2|\}$. Consequently,
the minimum of $r_\pi$ over all partitions of $A$ into 2 parts is
$\lceil \frac{m}{2} \rceil$.
\item If $k=3$ then $r_\pi= \max \{|A_1|,|A_2|,|A_3|, \lfloor
\frac{m}{2} \rfloor \}$. Consequently, the minimum of $r_\pi$ over
all partitions of $A$ into 3 parts is $\lfloor \frac{m}{2}
\rfloor$.
\end{itemize}
\end{proposition}

\begin{proof}

In each case, we determine the maximum of $s(\Delta)$ over all
families $\Delta$ that are feasible for $\pi$.

For $k=1$, a feasible family consists of at most $|A_1|=m$ copies
of $\{1\}$, and therefore $\max s(\Delta)=m$.

A feasible family for $k=2$ cannot contain two sets, $H_i$ and
$H_j$, such that $1 \notin H_i$ and $2 \notin H_j$, because such
sets would be disjoint. Hence, for any feasible family $\Delta$,
either all sets contain 1 or all of them contain 2. Therefore,
$s(\Delta) \le \max \{|A_1|,|A_2|\}$. On the other hand, feasible
families of size $|A_1|$, $|A_2|$ trivially exist.

Suppose $k=3$, and denote $$\beta_l = |A_l| \quad\mbox{for }
l=1,2,3.$$ We first show that $s(\Delta) \le \max \{ \beta_1,
\beta_2, \beta_3, \lfloor \frac{m}{2} \rfloor \}$ for every
feasible family $\Delta$. If $\Delta$ contains some singleton
$\{l\}$, then all sets in $\Delta$ must contain $l$, and hence
$s(\Delta) \le \beta_l$. Otherwise, $\Delta$ consists of $s_{12}$
copies of $\{1,2\}$, $s_{13}$ copies of $\{1,3\}$, $s_{23}$ copies
of $\{2,3\}$, and $s_{123}$ copies of $\{1,2,3\}$, for some non
negative integers $s_{12}, s_{13}, s_{23}, s_{123}$. We have the
following inequalities: $$s_{12}+s_{13}+s_{123} \le \beta_1,$$
$$s_{12}+s_{23}+s_{123} \le \beta_2,$$ $$s_{13}+s_{23}+s_{123} \le
\beta_3.$$ Upon adding these inequalities we obtain
$$2(s_{12}+s_{13}+s_{23})+3s_{123} \le m,$$ which implies
$$s(\Delta)=s_{12}+s_{13}+s_{23}+s_{123} \le \lfloor \frac{m}{2}
\rfloor.$$ We show next that there exists a feasible family
$\Delta$ with $s(\Delta)= \max \{\beta_1, \beta_2, \beta_3,
\lfloor \frac{m}{2} \rfloor \}$. If this maximum is one of the
$\beta_l$'s, this is trivial. So assume that $\beta_l < \lfloor
\frac{m}{2} \rfloor$ for $l=1,2,3$. If $m$ is even then the family
$\Delta$ that consists of $$s_{12}=\frac{\beta_1 +\beta_2
-\beta_3}{2} \quad\mbox{copies of } \{1,2\},$$
$$s_{13}=\frac{\beta_1 + \beta_3 - \beta_2}{2} \quad\mbox{copies
of } \{1,3\},$$ $$s_{23}= \frac{\beta_2 + \beta_3 - \beta_1}{2}
\quad\mbox{copies of } \{2,3\},$$ is feasible (note that the
prescribed numbers are non negative because $\beta_l < \lfloor
\frac{m}{2} \rfloor$ for $l=1,2,3$, and they are integers because
$\beta_1 + \beta_2 + \beta_3 =m$ is even). The size of this family
is $s(\Delta)=s_{12}+s_{13}+s_{23}= \frac{m}{2}$. If $m$ is odd,
we make slight changes in the values of $s_{12},s_{13},s_{23}$: we
add $\frac{1}{2}$ to one of them and subtract $\frac{1}{2}$ from
the other two. In this way we get a family $\Delta$ with
$s(\Delta)= \lfloor \frac{m}{2} \rfloor$.\QED
\end{proof}

We see from Proposition~1 that if we use partitions into two parts
(entailing a communication complexity of 4), the best we can do in
terms of economic efficiency is $r_\pi = \lceil \frac{m}{2}
\rceil$, and this is achieved by partitioning $A$ into equal or
nearly equal parts. Allowing for three parts (and therefore a
communication complexity of 8) permits only a small gain in
$r_\pi$ (in fact, no gain at all when $m$ is even).

We will now state the two parts of our main result.
 \begin{theorem}
Let $\pi = \{A_1,...,A_k\}$ be a partition of $A$ into $k$ non
empty sets of maximum size $\beta(\pi)$. (That is, $\beta(\pi) =
\max \{|A_1|,...,|A_k|\}$.) Then $$r_\pi \le \beta(\pi) \cdot
\varphi(k),$$ where $$\varphi(k)=\max_{j=1,...,k} \min \{
j,\frac{k}{j} \}.$$

 \end{theorem}

 The proof of Theorem~3 is given in the following subsection.

Note that $$\varphi(k)\le\sqrt{k}.$$ In particular, if all sets in
$\pi$ have equal size $\frac{m}{k}$, we obtain the upper bound
$$r_\pi \le \frac{m}{\sqrt{k}}.$$

Now, consider the case when, for some non negative integer $q$, we
have $$k=q^2+q+1, \quad\mbox{and} \eqno{(5.3)}$$ $$|A_i|=q+1
\quad\mbox{for } i=1,...,k. \eqno{(5.4)}$$ In this case
$$\varphi(k)= \frac{q^2+q+1}{q+1},$$ and hence the upper bound of
Theorem~3 takes the form $$r_\pi \le k.$$ The second part of our
main result implies that in infinitely many of these cases this
upper bound is tight.

 \begin{theorem}
Let $\pi = \{A_1,...,A_k\}$ be a partition that satisfies (5.3)
and (5.4) for some $q$ which is either 0 or 1 or of the form $p^l$
where $p$ is a prime number and $l$ is a positive integer. Then
$$r_\pi = k.$$

 \end{theorem}

We prove Theorems 3 and 4 in the following subsection.

\subsection{Proofs of Theorems~3 and 4}
We begin with some preparations. Let $\Delta=(H_i)_{i=1}^s$ be a
family of (not necessarily distinct) subsets of $\{1,...,k\}$. A
vector of non negative  numbers $\delta=(\delta_i)_{i=1}^s$ is
called a {\bf
 semi balanced}\footnote{This concept is equivalent to what is
 called a fractional matching in combinatorics. We chose the term
 semi balanced, because balanced vectors, defined by requiring
 equality instead of weak inequality, are a familiar concept in
 game theory (see e.g. \cite{Shapley67}).}
 vector for $\Delta$ if for every $l \in \{1,...,k\}$,
$$\sum_{i: l\in H_i}\delta_i\le 1.$$%
\begin{proposition}
Let $\Delta=(H_i)_{i=1}^s$ be a family of (not necessarily
distinct) subsets of $\{1,...,k\}$ such that $H_i\cap
H_j\ne\emptyset$ for every $1\le i,j\le s$. Let
$\delta=(\delta_i)_{i=1}^s$ be a semi balanced vector for
$\Delta$. Then $$\sum_{i=1}^s\delta_i\le\varphi(k),$$ where

$$\varphi(k)=\max_{j=1,...,k}\min \{j,\frac{k}{j}\}.$$
\end{proposition}
\begin{proof}

Assume without loss of generality that $h=|H_1|$ is the minimal
number of elements in a member of $\Delta$. The proposition will
be proved if we prove the following two claims:

 \noindent{\bf Claim 1:}
$\sum_{i=1}^s\delta_i\le h.$

\noindent{\bf Claim 2:} $\sum_{i=1}^s\delta_i\le\frac{k}{h}.$

\noindent{\bf Proof of Claim 1:}

Let $z=\sum_{l\in H_1}\sum_{i: l\in H_i}\delta_i$. As every $H_i$
intersects $H_1$, every $\delta_i$ appears in $z$ at least once.
Therefore, $z \ge \sum_{i=1}^s\delta_i$. Because $\delta$ is
 semi balanced, $\sum_{i: l\in H_i} \delta_i \le 1$ for every $l$, and in particular
for $l \in H_1$. Hence, $z\le \sum_{l\in H_1} 1=h.$ \QED

\noindent{\bf Proof of Claim 2:}

Let $w= \sum_{l=1}^k \sum_{i:l \in H_i} \delta_i$. Every
$\delta_i$ appears in $w$ exactly $|H_i|$ times. Since $|H_i| \ge
h$ for every $i$, we have $w \ge h \sum_{i=1}^s \delta_i$. On the
other hand, as in the proof of Claim~1, we obtain $w \le
\sum_{l=1}^k 1=k$. Combining the two inequalities, we get
$\sum_{i=1}^s \delta_i \le \frac{k}{h}$. \QED

 Therefore,
 $$\sum_{i=1}^s\delta_i\le
\min \{h,\frac{k}{h}\}\le\varphi(k).$$

 \QED

\end{proof}

We are now ready for the proof of Theorem 3:

\noindent{\bf Proof of Theorem 3:}

Let $\pi = \{A_1,...,A_k\}$ be a partition of $A$ into $k$ non
empty sets of maximum size $\beta(\pi)$. We have to prove that
$r_\pi \le \beta(\pi) \cdot \varphi(k)$. By Theorem~2, it suffices
to show that for every feasible family $\Delta$ for $\pi$, we have
$$s(\Delta) \le \beta(\pi) \cdot \varphi(k).$$ Let
$\Delta=(H_i)_{i=1}^s$ be such a family. Consider the vector
$\delta = (\delta_i)_{i=1}^s$ with $$\delta_i =
\frac{1}{\beta(\pi)}, \quad i=1,...,s.$$ By the second condition
of feasibility, this vector is semi balanced. Hence we may apply
Proposition~2 and conclude that $$\sum_{i=1}^s \delta_i \le
\varphi(k),$$ or equivalently, $$\frac{s}{\beta(\pi)} \le
\varphi(k),$$ as required. \QED

In order to prove Theorem~4 we invoke a result about finite
geometries (see e.g. \cite{Demb}). A {\bf finite projective plane}
of order $q$ is a system consisting of a set $\Pi$ of points and a
set $\Lambda$ of lines (in this abstract setting, a line is just a
set of points, i.e., $L \subseteq \Pi$ for every $L \in \Lambda$),
satisfying the following conditions:
\begin{itemize}
\item $|\Pi|=|\Lambda|=q^2+q+1$.
\item Every point is incident to $q+1$ lines and every line
contains $q+1$ points.
\item There is exactly one line containing any two points, and
there is exactly one point common to any two lines.
\end{itemize}
Such a system does not exist for every $q$. However, it trivially
exists for $q=0$ (a single point) and for $q=1$ (a triangle) and
it is known to exist for every $q$ of the form $q=p^l$, where $p$
is a prime number and $l$ is a positive integer. The first non
trivial example, corresponding to $q=2$, is called the {\bf Fano
plane}: $$\Pi= \{1,2,3,4,5,6,7\},$$ $$\Lambda=
\{124,235,346,457,561,672,713\}.$$

\noindent {\bf Proof of Theorem~4:}

Let $\pi = \{A_1,...,A_k\}$ be a partition that satisfies (5.3)
and (5.4) for some $q$ which is either 0 or 1 or of the form $p^l$
where $p$ is a prime number and $l$ is a positive integer. As
$r_\pi \le k$ follows from Theorem~3 (see the discussion preceding
the statement of Theorem~4), we need to prove only that $r_\pi \ge
k$. By Theorem~2, it suffices to show that there exists a family
$\Delta$ with $s(\Delta)=k$ which is feasible for $\pi$. Such a
family is given by the system of lines of a projective plane of
order $q$, when the points are identified with $1,...,k$. \QED

\subsection{More on the ranking of equilibria}

The tradeoff between communication complexity and economic
efficiency, as delineated above, may be made concrete by the
following scenario. Suppose that a set $A$ of $m$ goods is given,
and we are in a position to recommend to the potential buyers an
equilibrium strategy. Assume further that a certain level $M$ of
communication complexity is considered the maximum acceptable
level. If we are going to recommend a partition-based equilibrium
$f^\pi$, then the number of parts in $\pi$ should be at most $k=
\lfloor \log_2 M \rfloor$. From the viewpoint of economic
efficiency, we would like to choose such a partition $\pi$ with
$r_\pi$ as low as possible. Which partition should it be?

According to Theorem~3, we obtain the lowest guarantee on $r_\pi$
by making the maximum size of a part in $\pi$ as small as
possible, which means splitting $A$ into $k$ equal (or nearly
equal, depending on divisibility) parts. This leads to the
question whether, for given $m$ and $k$, the lowest value of
$r_\pi$ itself (not of our upper bound) over all partitions $\pi$
of $A$ into $k$ parts is achieved at an equi-partition, i.e., a
partition $\pi =\{A_1,...,A_k\}$ such that $\lfloor \frac{m}{k}
\rfloor \le |A_i| \le \lceil \frac{m}{k} \rceil$, $i=1,...,k$.

While Proposition~1 gives an affirmative answer for $k=1,2,3$, it
turns out, somewhat surprisingly, that this is not always the
case. This is shown in the following example.

\noindent {\bf Example~3}

Let $m=21$ and $k=7$. If $\pi$ is an equi-partition of the 21
goods into 7 triples then, by Theorem~4, $r_\pi =7$. Consider now
a partition $\pi'=\{A_1,...,A_7\}$ in which $$|A_1|=2, |A_2|=4,
|A_3|= \cdots = |A_7|=3.$$ We claim that $r_{\pi'} \le 6$.

In order to prove this, it suffices to show that there exists no
feasible family of 7 sets for $\pi'$. Suppose, for the sake of
contradiction, that $\Delta=(H_i)_{i=1}^7$ is such a family. Let
$H_i$ be an arbitrary set in $\Delta$. It follows from the second
condition of feasibility that if $H_i$ contains the element 1 then
it shares it with at most one other set in $\Delta$. Similarly, if
$H_i$ contains the element 2 then it shares it with at most three
other sets in $\Delta$. For $l=3,...,7$, if $H_i$ contains the
element $l$ then it shares it with at most two other sets in
$\Delta$. This implies that $H_i$ must contain at least three
elements (because it must share an element with every other set,
and $3+2<6$). Moreover, if $H_i$ contains exactly three elements
and one of them is 1, then it also contains 2 (since $1+2+2<6$).
On the other hand, we have $$\sum_{i=1}^7 |H_i| = \sum_{i=1}^7
\sum_{l \in H_i} 1= \sum_{l=1}^7 \sum_{i:l \in H_i} 1=
\sum_{l=1}^7 |\{i:l \in H_i\}| \le \sum_{l=1}^7 |A_l|=21.$$ Since
every $H_i$ has at least three elements, it follows that every
$H_i$ has exactly three elements, and all the weak inequalities
$|\{i:l \in H_i\}| \le |A_l|$ must in fact hold as equalities. In
particular, there exist two sets in $\Delta$, say $H_i$ and $H_j$,
that contain the element 1. By the above, they both contain 2 as
well. Let $l$ be the third element of $H_i$. Then among the
remaining five sets in $\Delta$, the set $H_i$ shares the element
1 with none of them, it shares the element 2 with two of them, and
the element $l$ with at most two of them. This contradicts the
fact that $H_i$ intersects every other set in $\Delta$.

It can be checked that in fact $r_{\pi'}=6$ and this is the lowest
achievable value among all partitions of 21 goods into 7 sets. We
omit the detailed verification of this.

The tradeoff between communication complexity and economic
efficiency was quantitatively analyzed above only for
partition-based equilibria. It is natural to ask whether it is
possible to beat this tradeoff using the more general bundling
equilibria. The answer is, in a sense made precise below:
sometimes yes, but not by much.

\noindent {\bf Example~4}

Assume that the number of goods $m$ is even, and let the set of
goods $A$ be partitioned into two equal parts $B$ and $C$.
Consider $\Sigma \subseteq 2^A$ defined by $$\Sigma = \{ D
\subseteq A: |D \cap B|=|D \cap C| \}.$$ It is easy to check that
$\Sigma$ is a quasi field, and hence it induces a bundling
equilibrium. The communication complexity is
$$|\Sigma|=\sum_{j=0}^{m/2} {m/2 \choose j}^2=\sum_{j=0}^{m/2}
{m/2 \choose j}{m/2 \choose m/2 -j}= {m \choose m/2}.$$

We claim that $r_\Sigma =2$. That $r_\Sigma \ge 2$ can be seen by
taking two buyers with valuations $w_B$ and $w_C$, respectively.
To see that $r_\Sigma \le 2$, suppose that $v$ is a profile of
valuations for a set of buyers $N$, and let $\gamma$ be an optimal
allocation. Split the set $N$ into the following two sets: $$N_B=
\{i \in N:|\gamma_i \cap B| \ge |\gamma_i \cap C| \},$$ $$N_C= \{i
\in N:|\gamma_i \cap B| < |\gamma_i \cap C| \}.$$ Note that the
sets of goods $\gamma_i, i \in N_B$, can be expanded to pairwise
disjoint sets of goods that belong to $\Sigma$. In other words,
there exists a $\Sigma$-allocation $\xi$ such that $\gamma_i
\subseteq \xi_i$ for every $i \in N_B$. Similarly, there exists a
$\Sigma$-allocation $\eta$ such that $\gamma_i \subseteq \eta_i$
for every $i \in N_C$. Hence $$S_{max}(v)= \sum_{i \in N}
v_i(\gamma_i)= \sum_{i \in N_B} v_i(\gamma_i) + \sum_{i \in N_C}
v_i(\gamma_i) \le \sum_{i \in N} v_i(\xi_i) + \sum_{i \in N}
v_i(\eta_i) \le 2S_\Sigma(v).$$ Thus, $r_\Sigma \le 2$.

We claim further that if a partition $\pi$ of $A$ satisfies $r_\pi
\le 2$ then $|\Sigma_\pi| \ge 2^{m-2}$. Indeed, suppose $\pi =
\{A_1,...,A_k\}$. It is easy to find a feasible family of 3 sets
for $\pi$ if one of the $A_l$'s has three or more elements, or if
three of the $A_l$'s have two elements each. Therefore, $r_\pi \le
2$ implies that at most two of the sets $A_1,...,A_k$ have two
elements and the rest are singletons. Thus $k \ge m-2$ and
$|\Sigma_\pi| \ge 2^{m-2}$.

Since ${m \choose m/2} < 2^{m-2}$ for all even $m \ge 10$, we have
the following conclusion: If $m \ge 10$ then every partition-based
equilibrium that matches the economic efficiency of $f^\Sigma$ has
a higher communication complexity than $f^\Sigma$. In other words,
the quasi field $\Sigma$ offers an efficiency/complexity
combination that cannot be achieved or improved upon (in the
Pareto sense) by any field.

The above example notwithstanding, the efficiency/complexity
combinations which arise from arbitrary quasi fields are still
subject to a tradeoff that is not much better than for fields. This
is the content of our final remark.

\noindent {\bf Remark 2}
 Let $m=|A|$ and let $k$ be a positive integer. Any quasi field
$\Sigma \subseteq 2^A$ with $r_\Sigma \le \frac{m}{k}$ must
contain a partition of $A$ into $k$ non empty parts, and therefore
must satisfy $|\Sigma| \ge 2^k$.

 \noindent {\bf Proof:}

Let there be $m$ buyers, each with valuation $w_a$ for a distinct
$a \in A$. For this $v$ we have $S_{max}(v)=m$. If $r_\Sigma \le
\frac{m}{k}$ then we must have $S_\Sigma(v) \ge k$. Hence an
optimal $\Sigma$-allocation has to assign non empty bundles of
goods to at least $k$ buyers. Thus $\Sigma$ contains $k$ pairwise
disjoint non empty sets of goods, and therefore, being a quasi
field, also a partition of $A$ into $k$ non empty parts. \QED


\end{document}